\documentclass[]{aa}
\usepackage{psfig}
\begin{document}

\thesaurus{07(08.19.5; 11.19.1; 03.20.8)}

\title{Early-time spectroscopic observations of \\
SN~1998aq in NGC~3982}

\author{J.Vink\'o \inst{1}\inst{2} \and L.L.Kiss \inst{3} \and
J.Thomson \inst{4} \and G.F\H ur\'esz \inst{1} \and W.Lu \inst{4}
\and G.Kasz\'as \inst{1} \and Z.Balog \inst{1}}

\institute{Department of Optics \& Quantum Electronics, JATE University,
Szeged, POB 406, H-6701 Hungary \and
Research Group on Laser Physics of the Hungarian Academy of Sciences \and
Department of Experimental Physics, JATE University, Szeged, D\'om t\'er 9.
H-6720 Hungary \and
David Dunlap Observatory, University of Toronto, Box 360, Richmond Hill, 
Ontario L4C 4Y6 Canada}

\titlerunning{Spectroscopy of SN~1998aq}
\authorrunning{J.Vink\'o et al.}
\offprints{vinko@physx.u-szeged.hu}
\date{}

\maketitle
 
\begin{abstract}

Medium resolution spectroscopic observations of SN~1998aq made in April -
May, 1998, are presented. We confirm its Type~Ia classification based
on the prominent Si~II $\lambda$6355 line and other features similar
to SNe~Ia. No significant sign of hydrogen Balmer-alpha line was
seen in high-resolution spectra taken at 8 days before maximum.
Preliminary estimate of reddening and distance modulus, based on
the correlation  between the line depth ratio of two Si~II features 
around 6000 \AA~and $BVRI$ light curves of Type~Ia SNe 
(Riess et al., \cite{riess2}), is presented. 
Moreover, $H\alpha$ emission from the core of the Seyfert~2 
type host galaxy NGC~3982 has also been detected.

 
\end{abstract}

\section{Introduction}
The number of spectroscopic observations of supernovae increased quickly in
this decade, partly due to the advances of CCD-technique and the growing
number of supernova searching observational projects (see Filippenko,
\cite{filip} for a detailed review). In the first half of 1998, three SNe
with brightness of about $V{\approx}12$~mag (1998S, 1998aq and 1998bu) was
discovered together with many more fainter ones. In this paper we report
medium-resolution spectroscopic observations of SN~1998aq made shortly
before and after maximum light.

SN~1998aq was discovered by M.~Armstrong (Hurst et al., \cite{hurst}).
It has been classified as SN~Ia by Ayani \& Yamaoka (\cite{ayani}) who
reported prominent Si~II $\lambda$6355 and other S~II, Fe~II and Mg~II
absorption lines which made SN~1998aq similar to the ``prototype'' SN~Ia
SN~1994D. The expansion velocity was determined as about 11,000~km/s. 
They also
pointed out the absence of Na~D absorption due to probably small
interstellar reddening. Shortly later, Berlind \& Calkins (see Garnavich
et al., \cite{garnav}) reported the similarity to SN~1990N based on a
spectrum obtained at 1~week before maximum.

Another interesting property of SN~1998aq is that its host galaxy, 
NGC~3982 (PGC~37520, UGC~6918, IRAS~11538+5524),
has a Seyfert~2 type nucleus. This galaxy was a subject of a recent 
AGN-survey by Ho
et al. (\cite{ho}). To date, there is an indication that SNe in the host
galaxies of AGNs show higher concentration toward the galaxy core with
respect to SNe in normal galaxies (Petrosian \& Turatto, \cite{petros}). This
may give evidence on increased star formation rate in the proximity of AGN,
but the number of actually observed SNe in such systems is not large, so
more data would significantly improve the statistics.

\section{Observations}
\begin{figure*}
\leavevmode
\hbox{\psfig{figure=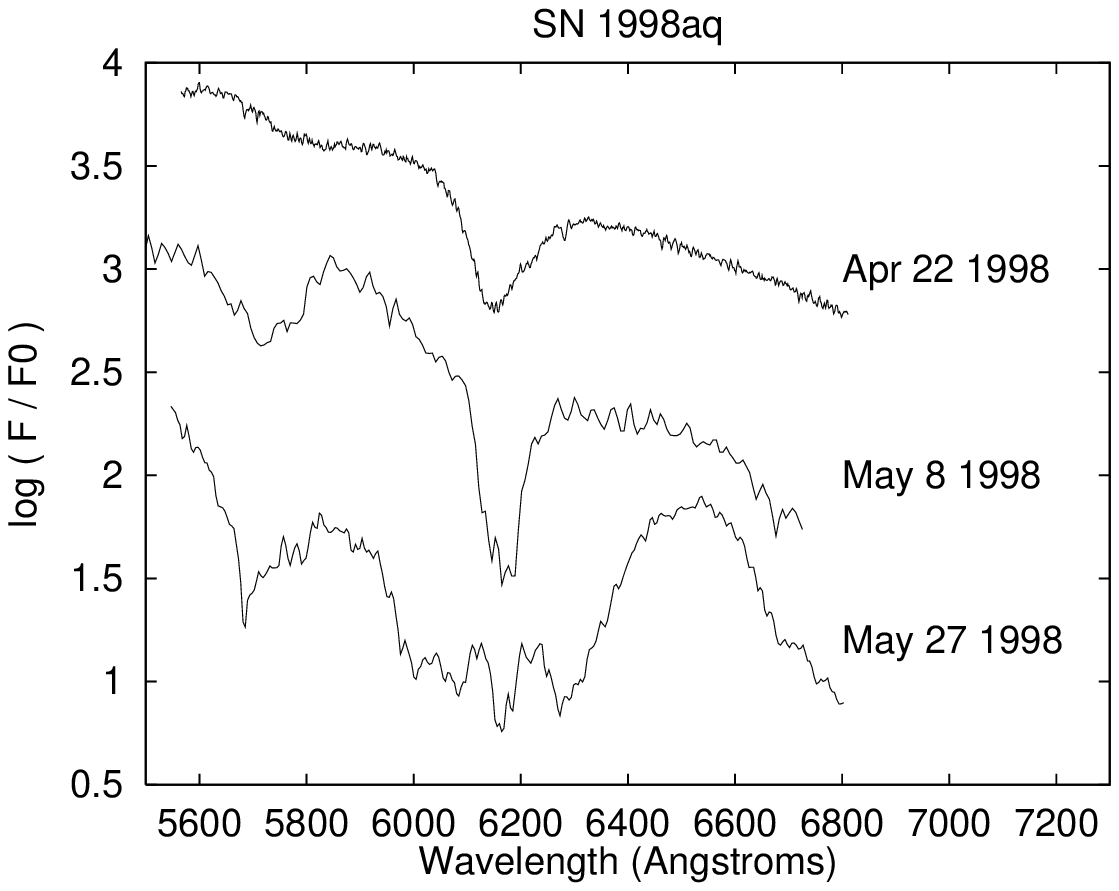,width=11cm,height=7cm}\hspace{1cm}
     \psfig{file=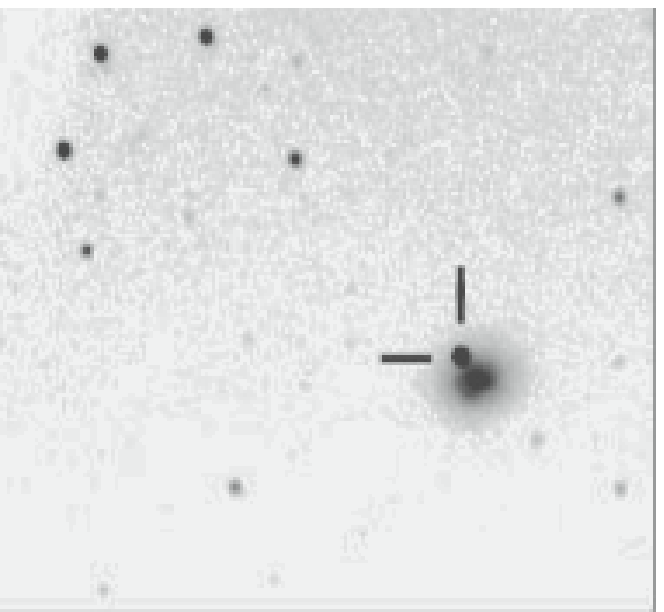,width=6cm,height=5.5cm}}
\caption{Observed spectra of SN~1998aq (left panel) and
CCD picture of NGC~3982 showing SN~1998aq (right panel).
The epochs of spectroscopic observations are indicated 
on the right side of each spectrum.}
\label{fig_1}
\end{figure*}

We made medium- and high-resolution spectroscopic observations of SN~1998aq
between April 22th and May 27th, 1998 at David Dunlap Observatory, Canada with
the 74" Cassegrain telescope. The gratings used were the 150~lines/mm (in
2nd order with an order-separation filter inserted) and the 1800~lines/mm
giving 1.3~\AA~ per pixel and 0.2~\AA~ per pixel resolution, respectively.
The medium-resolution spectra are presented in Fig.~\ref{fig_1} 
(left panel) where an
arbitrary vertical shift has been added to each spectrum for better 
visibility. The decrease of
the signal-to-noise ratio toward the later spectra was due to 
the faintening and the
increasing airmass of the object in May. The data were reduced by standard
{\it IRAF} routines. FeAr spectral lamp exposures were used for wavelength
calibration. Particular attention was payed to remove the background light
contamination due to the host galaxy (discussed below) and the night sky. An
unfiltered CCD-image showing SN~1998aq in NGC~3982 taken from downtown of
Szeged with a 11" Schmidt-Cassegrain telescope and ST-6 camera is presented
in the right panel of Fig.~\ref{fig_1}.

\begin{figure}
\leavevmode
\psfig{figure=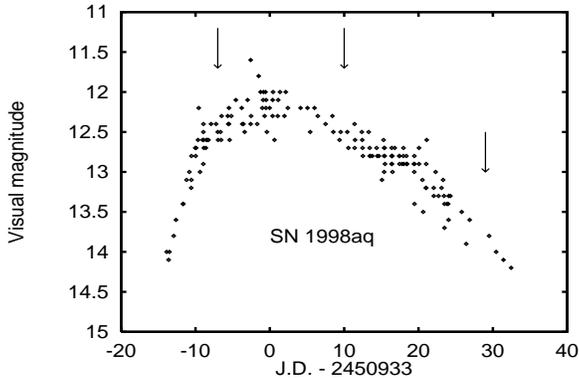,width=8cm,height=5cm}
\caption{Visual light curve of the SN1998aq based on the {\it VSNET}
data bank. Arrows indicate the epochs of spectroscopic observations.}
\label{fig_2}
\end{figure}

In order to determine the phase of our spectra relative to the light
curve of the SN, we collected all available visual observations of SN~1998aq
made by amateur astronomers, using
the public database of the {\it Variable Star Observers' Network
({\it VSNET})}. This light curve is plotted in  
Fig.~\ref{fig_2}. The typical uncertainty of the individual points 
is at least
$\pm$0.3 mag and the spread of the light curve goes up to 0.5 mag
at a given epoch, although the observers used mainly the same sequence
of comparison stars.
As a first approximation, we estimated the moment
of visual maximum light being JD 24,50933. Using this epoch we 
determined the approximate phase of our spectra as 
1~week before maximum, 9~days after maximum and 28~days after maximum,
respectively (see also Table~1). 

\section{Results and discussion}

\subsection{Spectral characteristics}
Although the spectra presented in Fig.~1 clearly have inadequate wavelength
coverage for a detailed comparison with other SNe spectra, 
some basic properties of SNe Type Ia can be recognized in these data.
The most prominent feature in the two earlier spectra is the Si~II
$\lambda$6355 absorption line as noted by other observers. This is the
characteristic feature of SNe~Ia (e.g. Filippenko, \cite{filip}). 
On the pre-maximum spectrum, the line profile is asymmetric and has a slight
P~Cyg-type ``bump'' toward longer wavelengths, similarly to SN~1994D
(Patat et al., \cite{patat}). Later, this absorption line deepens
and broadens significantly at about 1~month
post-maximum, which is also similar to the spectral behaviour of SN~1994D in
this wavelength regime, although Berlind \& Calkins reported ``unusually
shallow Type~Ia features'' (see Garnavich et al., \cite{garnav}).
The other Si II absorption trough at 5700 \AA~becomes
stronger as the SN gets older, but it also becomes blended from
its blue side.
Moreover, the broad emission bump at $\lambda$6500 at $20-40$~days
post-maximum, which is probably due to Fe~II and $[$Fe~III$]$ (Filippenko,
\cite{filip}) is also reproduced well on the third spectrum. These observed
features indicate that SN~1998aq closely resembles to a ``prototype'' SN~Ia
in the $5500-6700$~\AA~spectral interval. 

\subsection{Radial velocities}
We have derived velocities of the expanding gas measuring
the Doppler-shift of the line core of the $\lambda 6355$ Si~II line 
(Table~1). Such ``line-core'' velocities have been 
presented for a number of other SNe~Ia by Patat et al.
(\cite{patat}, see their Fig.10). According to that diagram, 
the velocities of SN~1998aq agree well with those of SN~1994D and
SN~1989B. However, as it was also noted by Patat et al.
(\cite{patat}), the velocities derived from strong lines, such
as Si~II $\lambda 6355$, are ambiguous, especially at later
phases, because these lines are formed over a considerably
large velocity range. It would be interesting to derive 
bisector velocities of both observed and synthesized SN spectra
to reveal the effect of velocity gradients, as it was recently 
done e.g. for Cepheids (Butler et al., \cite{butler}). 
The spectra presented in this paper do not have the 
necessary signal-to-noise and phase coverage for such purpose.

\begin{table}
\label{tabl_1}
\begin{center}
\caption{Phases of spectroscopic observations relative to the
moment of visual maximum, JD 24,50933 and  
radial velocities determined from the Si~II line (in the
host galaxy rest frame, $v_{gal}=1100$~km/s).}
\begin{tabular}{ccc}
\hline\noalign{\smallskip}
J.D. & $t$ & $v_{rad}$ \\
 & (days) & ($10^3$ km/s) \\
\noalign{\smallskip}\hline\noalign{\smallskip}
2450926 & $-7$ & $-10.8$ \\
2450942 & $+9$ & $-10.1$ \\
2450961 & $+28$ & $-10.0$ \\
\noalign{\smallskip}\hline
\end{tabular}
\end{center}
\end{table}

\subsection{Preliminary reddening and distance estimates}
A method for obtaining ``snapshot'' distances to SNe~Ia has been
developed very recently by Riess et al. (\cite{riess2}). 
The idea is the following: one can calculate the distance of the
SN by comparing a single $BV$ or $BVRI$ photometric measurement 
with a calibrated template SN Ia light curve (describing the 
absolute magnitude of an ``ideal'' SN Ia as a function of 
time) if the phase of the photometric data and 
the ``light-curve parameter'' $\Delta$ (giving the magnitude 
difference between the maximum brightness of the observed and
the template SN) is known. A template SN~Ia light curve has 
been given by Riess et al. (\cite{riess}). 
For the determination of $\Delta$, a correlation is found
between $\Delta$ and the ratio of line depths of the Si II 
absorption lines at 5800 \AA~and 6150 \AA~(Nugent et al.,
\cite{nugent}; Riess et al., \cite{riess2}). 

\begin{figure}
\begin{center}
\leavevmode
\psfig{file=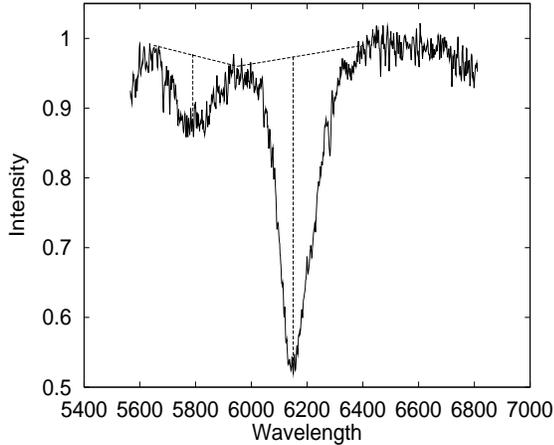,width=8cm,height=6cm}
\caption{Graphical representation of the method of
measuring line depth ratio of the two Si~II line.}
\end{center}
\label{fig_3}
\end{figure}

We tried to apply the method outlined above using the first spectrum,
obtained on April 22th. As the first step, we normalized the spectrum
to the continuum by fitting a smoothly varying Chebyshev-function
to the highest flux levels of the spectrum in order to correct for
the steep decline of the intensity toward longer wavelengths. 
After that, we measured the line depths of the Si~II troughs as
shown in Fig.~\ref{fig_3} (following the prescription given
by Nugent et al., \cite{nugent}). 
The ratio, $R$, of these depths were then calculated, resulting in 
$R$(Si~II) = $d$(5800\AA)$/d$(6150\AA)~=~$0.22 \pm 0.02$.
Using the linear relationship between $R$(Si~II) and $\Delta$
at $t=-7$ days relative to maximum light (Riess et al.,
\cite{riess2}), $\Delta = 0.06 \pm 0.03$ was derived.
The low value of $\Delta$ means that 
the light curve of SN~1998aq {\it may} not deviate largely from
the template SN~Ia light curve (but it should, of course, be
proven by extensive photometry of the SN, which, unfortunately,
was not available for us during the preparation of the manuscript).

Strictly speaking, the calibration of $R$(Si~II) $vs~\Delta$ uses
the phase $t$ (in days) relative to the $B-maximum$ of the SN light curve
(Riess et al., \cite{riess2}). Because of the same reason
as above, we could only estimate the phase of our spectra using
the visual light curve plotted in Fig.~\ref{fig_2}. 
However, because of the low value of $\Delta$, this approximation
probably does not introduce large errors.  
Indeed, assuming that the phase of the first spectrum is 
$t=-6$ days, $\Delta=0.008$ could be
obtained which would further reduce the expected difference 
between the ``real'' light curve and the template light curve.
On the other hand, $t=-8$ days is improbable, because the
maximum light in $B$ occurs earlier than in $V$.
Again, this question should be re-investigated using  
calibrated long-term photometry of SN~1998aq.

As far as the available photometry is concerned, there are some
$BV$ measurements at the earlier phases of SN~1998aq published
in {\it IAU} Circulars. Although the accuracy
of the photometric data published in {\it IAU} Circulars is quite
variable and sometimes inferior, but, as above, it is the only 
source of publicly available calibrated photometry of SN~1998aq 
at the date of the preparation of this paper. 

We have collected $V_{obs}=12.67$ and $(B-V)_{obs}=0.02$ 
magnitudes, observed at April 20.904 UT (Hanzl \& Caton, \cite{hanzl}). 
These measurements were obtained with an ST-7 CCD-camera attached to a
40 cm Cassegrain telescope, according to one of the authors' (Hanzl)
description. The accuracy of these data should be much higher than 
the amateur visual light estimates showed in Fig.~\ref{fig_2} (which
is used only for estimating the phases of our spectroscopic 
measurements). Hanzl (\cite{hanzl2}) gives error estimates of 
his measurements 
in a follow-up publication, and typical values are
$\delta V = 0.01$ and $\delta (B-V) = 0.02 - 0.03$ mag. 
However, the anonymous referee of the present paper argued
that his own high-precision photometry gave
$B-V = -0.17$ on April 20. This means that the $B-V$ colour of
SN~1998aq may be much bluer than the single measurement of 
Hanzl \& Caton (\cite{hanzl}) indicates. 
We cannot discuss this discrepancy further, because it
is based on yet unpublished measurements, except to take it into account
in estimating the errors (see below).

Adopting $\tau = -8$ days as the phase of these photometric data,
the following relations have been applied to estimate the 
reddening and the distance modulus 
(Riess et al., \cite{riess}):

\begin{equation}
E(B-V) ~=~(B-V)_{obs} - (B-V)_0 - R_{B-V}(\tau) \Delta
\end{equation}
and
\begin{equation} 
{\mu}_0 ~=~V_{obs} - M_V(\tau) - R_V(\tau) \Delta - 3.1 E(B-V).
\end{equation}

We have adopted $R_V(\tau = -8) = 1.259$ and $R_{B-V}(\tau = -8) = 0.494$ 
from Table 2 of Riess et al. (\cite{riess}) and the standard value
of the galactic extinction law ($A_V / E(B-V) = 3.1$). 
This last assumption means that
the reddening ratio in the Milky Way can be used to describe the reddening
in distant galaxies, which was also favored by Riess et al. 
(\cite{riess}).

\begin{figure}
\begin{center}
\leavevmode
\psfig{figure=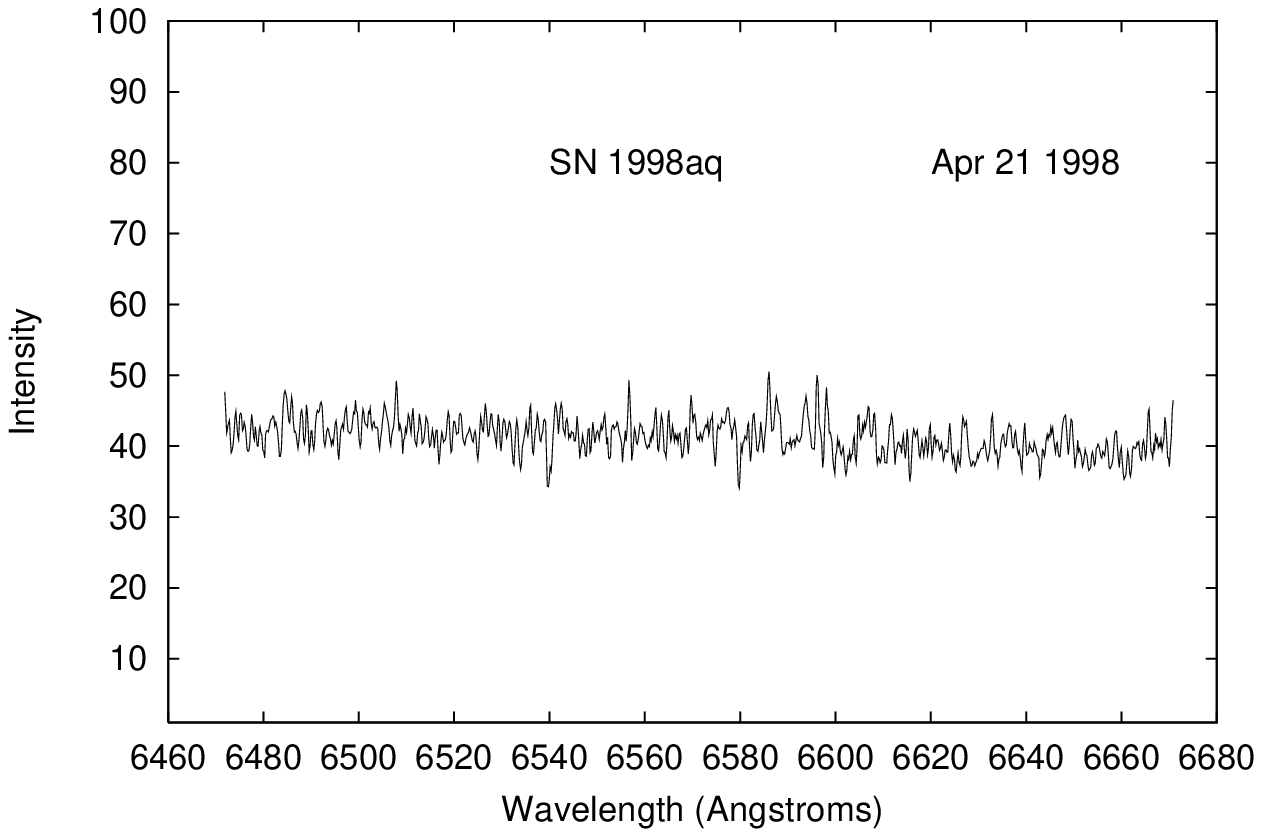,width=7cm,height=5cm}
\psfig{figure=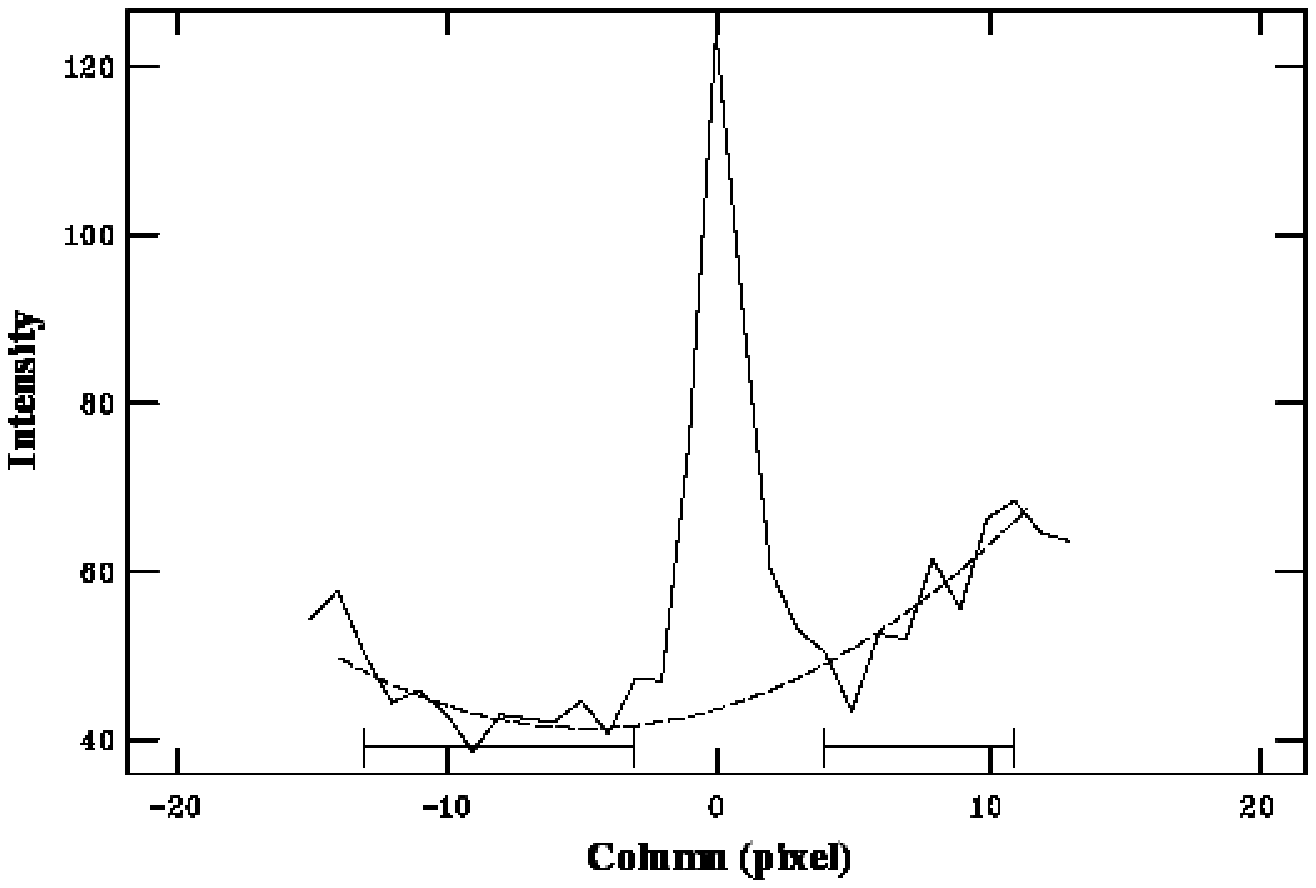,width=7cm,height=5cm}
\caption{\it{Up}: High resolution spectrum of SN~1998aq in the $H\alpha$ region 
obtained at 1 week before maximum. No significant sign of the hydrogen 
Balmer-alpha line is present. \it{Bottom}: Subtraction of background
contamination from the spectrum.}
\label{fig_4}
\end{center}
\end{figure}

We tried to estimate the colour excess $E(B-V)$ in two ways.
First, we adopted $(B-V)_{obs} = 0.02 \pm 0.03$ (Hanzl \& Caton, 
\cite{hanzl}), $(B-V)_0 = -0.244$ (Riess et al., \cite{riess}) 
and used Eq.(1) to get $E(B-V) = 0.23 \pm 0.04$ mag.
Second, we used the $COBE/IRAS$ All-Sky Reddening Map published
very recently by Schlegel et al. (\cite{schlegel}). This gives
the reddening toward a specified direction based on a calibration
between colour excess and the infrared flux at 100$\mu$. The query in
the direction of SN~1998aq resulted in $E(B-V) = 0.014 \pm 0.01$ mag. 
This low reddening value supports the suspicion that the observed
$(B-V)$ color of SN~1998aq around maximum was actually bluer than 
the only one published measurement of Hanzl (Hanzl \& Caton, \cite{hanzl}).
On the other hand, the host galaxy NGC~3982 has an active
nucleus of a Seyfert~2 type (see the next section), 
thus, higher dust concentration within the host galaxy 
might not be unrealistic. If this were the case, then the reddening
of SN~1998aq would be mainly due to dust absorption in its host
galaxy, rather than that in the Milky Way. Finally, we can consider 
$E(B-V) = 0.13 \pm 0.11$ mag as the unweighted average of the two 
data above, emphasizing the urgent need for published precise 
photometric measurements to solve this important and interesting problem.  

In order to derive the distance modulus via Eq.(2), the following
data were adopted: $V_{obs}=12.67 \pm 0.01$ mag (Hanzl \& Caton, 
\cite{hanzl}), $M_V(\tau = -8) = -18.693$ (Riess et al., 
\cite{riess}) and $R_V \Delta = 0.076$ mag (from the values above). 
The uncertainty of $M_V$ and $R_V \Delta$ was estimated as being
$\pm 0.2$ and $\pm 0.01$ mag, respectively, allowing $\pm 1$ day 
error in the epoch of the spectroscopic measurement. From the
reddening estimated above, the total absorption is
$A_V = 3.1 E(B-V) = 0.40 \pm 0.34$ with stronger probability that
the actual value is lower than this estimate. 

Substituting these values into Eq.(2), 
we get the extinction-free distance modulus as $\mu_0 = 30.89 \pm 0.56$ 
mag. The distance of the SN, corrected for interstellar absorption, is
$d ~=~ 15.1 \pm 4.4$ Mpc. However, it is stressed, that this
value can be considered only preliminary, which need
further confirmation based on much more extensive datasets. The
relatively large uncertainty of the distance reflects mainly
the lack of precise photometric information on this object.

\subsection{Hydrogen lines}
There have been controversial evidence of hydrogen Balmer-lines in the
spectra of SNe~Ia presented in the literature (see Filippenko, \cite{filip}
for review). In order to study the presence/absence of any $H\alpha$
feature in SN~1998aq, two consecutive spectra with higher resolution was
obtained on April 21th, about 8 days before maximum. The 
contamination of the light of the host galaxy was removed by fitting a
parabolic function outside the profile of the SN spectrum 
(Fig.~\ref{fig_4}, bottom panel), similarly to Della Valle 
et al. (\cite{della}). As it can be seen in the upper panel of
Fig.~\ref{fig_4}, no convincing detection of $H\alpha$ could be made. 

\begin{figure}
\begin{center}
\leavevmode
\psfig{figure=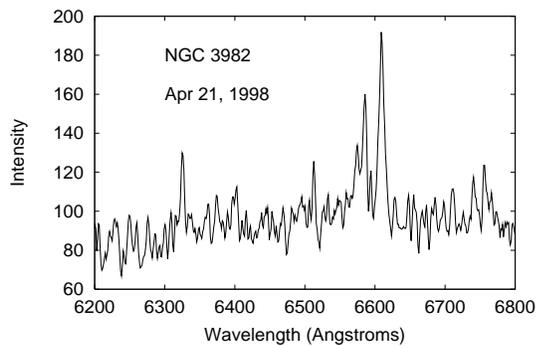,width=7cm}
\caption{Observed spectrum of the core of the host galaxy NGC~3982. The
emission pattern around 6600 \AA~is due to $H\alpha$, $[$N~II$]$ 
and $[$S~II$]$.}
\label{fig_5}
\end{center}
\end{figure}

As it has been mentioned above, the host galaxy, NGC~3982, 
has a Seyfert~2 type nucleus showing $H\alpha$ and some other
forbidden lines in emission (Ho et al., \cite{ho}). 
We have obtained
one spectrum of the core region of NGC~3982 which is plotted in
Fig.~\ref{fig_5}. The emission structure around 6600 \AA~consisting of 
$H\alpha$,
$[$N~II$]$ and $[$S~II$]$ is clearly detected, although the profile shapes
and relative strengths are different from those presented by Ho et al.
(\cite{ho}). This is probably due to our lower quality spectra and the lack
of sophisticated starlight-subtraction such as that applied by 
Ho et al. (\cite{ho}).

\section{Summary}
The summary of the results presented in this paper is as follows:

1. We obtained medium-resolution spectra around 6000~\AA~of SN~1998aq
before and after maximum. Based on the spectral features and the time
evolution of the spectrum, the classification of Type~Ia is confirmed. 
The decreasing expansion velocities are in agreement with other SN~Ia
velocities.

2. We applied the ``snapshot distance estimate'' method developed
by Nugent et al. (\cite{nugent}) and Riess et al. 
(\cite{riess2}) to the spectrum of SN~1998aq taken on April 22th.
The analysis resulted in $E(B-V) = 0.13 \pm  0.11$ mag and
$\mu_0 = 30.89 \pm 0.56$ mag as the value of the reddening and
the extinction-free distance modulus of SN~1998aq, respectively.
It is probable that the correct distance modulus is larger than
the mean value presented above, due to uncertainties in the
reddening.

3. High-resolution spectra obtained 8 days before maximum do not contain
any convincing feature that could be attributed to $H\alpha$. This is also
in agreement with the previous lack of detection of hydrogen lines in the
spectra of SNe~Ia. On the other hand, there is a pronounced 
$H\alpha$ emission emerging 
from the Seyfert~2 type nucleus of the host galaxy NGC~3982 which
was necessary to be taken into account during the reduction of
the high-resolution spectrum of SN~1998aq.

\begin{acknowledgements}
This research was supported by Hungarian OTKA Grants \#F022249,
\#T022259 and Foundation for Hungarian Education and Science 
(AMFK). The NASA ADS Abstract Service, the Canadian Astronomy Data
Center and the Variable Star Network (VSNET) was used to access
data and references. The $COBE/IRAS$ All-Sky Reddening Map has been
downloaded from the URL {http://astro.berkeley.edu/davis/dust}.
The availability of these services are gratefully acknowledged.
\end{acknowledgements}

\end{document}